# ENHANCED NONLINEAR INTERACTION OF POLARITONS VIA EXCITONIC RYDBERG STATES IN MONOLAYER $WSe_2$

Jie Gu[1,2], Valentin Walther[3], Lutz Waldecker[4], Daniel Rhodes[5], Archana Raja[6], James C. Hone[5], Tony F. Heinz[4], Stéphane Kéna-Cohen[7], Thomas Pohl[3], Vinod M. Menon[1,2,*]

[1]*Department of Physics, City College of New York, 160 Convent Ave., New York, NY 10031, USA*

[2]*Department of Physics, Graduate Center of the City University of New York (CUNY), 365 Fifth Ave. New York, NY 10016, USA*

[3]*Department of Physics and Astronomy, Aarhus University, Ny Munkegade 120, DK 8000 Aarhus, Denmark*

[4]*Department of Applied Physics, Stanford University, Stanford, California, 94305 USA*

[5]*Department of Mechanical Engineering, Columbia University, New York, NY 10027 USA*

[6]*Molecular Foundry, Lawrence Berkeley National Laboratory, Berkeley, CA, 94720 USA*

[7]*Department of Engineering Physics, École Polytechnique de Montréal, Montréal, Quebec, Canada*

---

[*] Email: vmenon@ccny.cuny.edu

**Strong optical nonlinearities play a central role in realizing quantum photonic technologies. In solid state systems, exciton-polaritons, which result from the hybridization of material excitations and cavity photons, are an attractive candidate to realize such nonlinearities. Here, the interaction between excitons forms the basis of the polaritonic nonlinearity. While the interaction between ground state excitons generates a notable optical nonlinearity[1–5], the strength of such ground state interactions is generally not sufficient to reach the regime of quantum nonlinear optics and strong single-polariton interactions. Excited states, however, feature enhanced interactions and therefore hold promise for accessing the quantum domain of single-photon nonlinearities, as demonstrated with high-lying Rydberg states of cold atomic systems. Excitons in excited states have recently been observed in monolayer transition metal dichalcogenides[6–10]. Here we demonstrate the formation of exciton-polaritons using the first excited excitonic state in monolayer tungsten diselenide ($WSe_2$) embedded in a microcavity. Owing to the larger exciton size compared to their ground state counterpart, the realized polaritons exhibit an enhanced nonlinear response by more than an order of magnitude, as evidenced through a modification of the cavity Rabi splitting. The demonstration of excited exciton-polaritons in two-dimensional semiconductors and their enhanced nonlinear response presents the first step towards the generation of strong photon interactions in solid state systems, a necessary building block for quantum photonic technologies.**

Exciton-polaritons, quasiparticles arising from the strong coupling between cavity photons and excitons in semiconductors, allow for the observation of exotic physical phenomena such as condensation[11–13], superfluidity[14], solitons[15] and quantized vortices[16], can be engineered to emulate systems such as atomic lattices for potential applications as quantum simulators[17], as well as for optoelectronic applications such as low energy switches[18], transistors[19] and interferometers[20]. This array of rich physical phenomena and the associated applications stem from the half-light half-matter make up of these quasiparticles. The photonic component lends the properties such as long-range propagation, small effective mass, and spatial coherence while the matter (exciton) component provides the interactions, spin-selectivity and nonlinearity. The strength of this interaction depends on the fraction of the excitonic component present in the polaritons. In inorganic semiconductors host Wannier-Mott excitons which can be described by a hydrogenic model, where the hole plays the role of the proton, this interaction strength is proportional to the exciton binding energy and the square of the exciton Bohr radius[3,21]. Systems such as monolayer transition metal dichalcogenides (TMDs) that have large ground state (1S) exciton binding energies and also reasonable exciton radius (1 nm) have therefore become attractive platforms for exploring polariton physics and devices at elevated temperatures[22–26]. However, even in TMDs the small exciton radius has been an impediment to realizing large exciton-exciton interaction strengths for the 1S excitons[27,28]. This suggests the use of excited-state excitons as a potential approach to enhance polariton interactions and optical nonlinearities[29] by exploiting their larger Bohr radius.

Excited states of excitons have been studied in variety of materials such as $CuCl_2$, GaAs, TMDs and halide perovskites[6–10,30–34]. Strong coupling of cavity photons to excited exciton states has been demonstrated in the GaAs and the perovskite systems recently[35,36]. However, the main motivation to go to excited states – to enhance the interaction strength – is yet to be convincingly demonstrated in any system. Here, we report the realization of 2S exciton polaritons in an archetypical two-dimensional TMD, $WSe_2$, embedded in a monolithic microcavity and more importantly demonstrate their enhanced interaction strength owing to their larger exciton radius. The interaction strength of 2S exciton-polaritons is found to be ~ 16 times larger than that of the 1S exciton-polaritons in similar TMD systems.

The sample structure is shown in Fig. 1a, where a 12-period silicon nitride ($SiN_x$)/ silicon dioxide ($SiO_2$) distributed Bragg reflector (DBR) was first grown on silicon substrate via plasma enhanced chemical vapor deposition. Three layers of hexagonal boron nitride (hBN) encapsulated $WSe_2$ were then transferred on top of the DBR followed by spin coating 250 nm Poly (methyl methacrylate) (PMMA). A 40 nm silver layer was deposited to form the top mirror of the cavity. The details of the device fabrication can be found in the Methods section. Three layers of $WSe_2$ were used in these experiments due to the weaker oscillator strength (5-10 times) of the excited exciton state (2S) compared to the ground state exciton (1S) in TMDs (Supplementary Section S2). Typically, the Rabi splitting for monolayer 1S exciton polaritons in TMDs is on the order of few tens of meV[37]. Since the total exciton-photon interaction strength is proportional to the square root of exciton density, more layers are necessary to achieve an observable Rabi splitting for the 2S states. Figure 1b is an optical microscope image showing the three layers of WSe$_2$ after transfer onto the bottom DBR. To confirm the observation of the 2S state, we performed a room temperature angle resolved white light reflection contrast measurement directly after the multilayer stack was transferred onto the bottom DBR. The reflection contrast is defined as 1-R$_{sample}$/R$_{ref}$, where R$_{sample}$ and R$_{ref}$ are reflected intensities from the sample area and a bare DBR area, respectively. Here a positive contrast indicates absorption in the sample. Both 1S exciton (1.655eV) and 2S excitonic (1.780eV) absorption features can be seen in Fig. 1c and their spectral positions are labeled by the white arrows. The parabolic features in Fig. 1c arise from the DBR side band.

Fourier space (*k*-space) imaging was carried out to determine the dispersion of the strongly coupled states. Fig. 2a-2c shows *k*-space white light reflection at different temperatures (140K, 77K, 15K). The white solid lines are fits simulated using transfer matrix method. The white dashed lines represent the bare 2S exciton energy ($E_{exc,2S}$) and bare cavity dispersions ($E_{cav}$). At 140 K (Fig. 2a), the 2S exciton energy lies below the cavity resonance with a positive detuning of 11 meV and does not show strong coupling. Here detuning is defined as $\delta = E_{cav} - E_{exc,2S}$ at the angle of zero degree. The exciton energy blue shifts with decreasing temperature (Supplementary Section S2). By lowering the temperature (Fig.2b and 2c), we can then blue shift the exciton energy across the cavity resonance leading to strong exciton-photon hybridization and clear anticrossing (Fig. 2d) between the upper and lower polariton branches with a measured Rabi splitting of $\Omega_0 \approx 7.7$ meV.

In order to investigate the nonlinear optical response of the excited exciton states, we measured the cavity reflection spectrum for varying intensities of the incident white light (See Methods for details on experiment). Upon increasing the intensity, the energy splitting between the polariton branches gradually decreases and eventually vanishes at the highest intensities of 250 mW/µm$^2$ as shown in Fig. 3. Similar behavior was observed using the 1S exciton ground state in WS$_2$ (See Supplementary Section S3). Increasing the intensity can cause both a closing of the Rabi splitting[38] as well as an overall blue shift[1,2] of both polariton branches, while the former clearly dominates the nonlinear response in the present experiments. We can understand and describe this behavior quantitatively in terms of an exciton blockade, whereby the interaction between excitons prevents their simultaneous generation within a blockade radius $R_{bl}$ and thereby inhibits the formation of polaritons within the distance $R_{bl}$. In close analogy to the emergence of optical nonlinearities in cold atomic Rydberg gases, this mechanism gives rise to a nonlocal optical nonlinearity with a nonlinear kernel (Supplementary Section S4)

$$\chi^{(3)}(r) = \frac{\Omega_0^4}{|\Gamma|^2 \Gamma} \frac{iU(r)}{\Gamma + iU(r)} \qquad (1)$$

where $U(r)$ is the interaction between two excitons at a distance $r$ and $\Gamma = \gamma - 2i\Delta$ is determined by the width $\gamma$ of the exciton resonance and $\Delta$, which is the energy difference between the incident light and the exciton resonance (Supplementary Section S4). This nonlocal form of the optical response describes both an exciton blockade within a blockade radius, $R_{bl}$, determined by $U(R_{bl}) = \Gamma$, as well as a cavity line shift that arises from exciton-exciton interactions outside of $R_{bl}$. Considering a situation in which the interaction predominantly leads to an exciton blockade, one can obtain a simple expression (Supplementary Section S4) for the resulting nonlinear effect on the Rabi splitting

$$\Omega_n = \Omega_0(1 - \pi R_{bl}^2 n/2) \qquad (2)$$

in terms of the blockade radius $R_{bl}$ and the polariton density $n$. Interestingly, eq. (2) affords a simple geometrical interpretation, indicating that the effective cavity coupling decreases directly with the number of blocked polaritons, as given by the product of the polariton density and the blockade area $\pi R_{bl}^2$.

Figure 4a shows the energy of both polariton branches for the 2S states at their avoided crossing as a function of the input intensity, which relates directly to the number of injected polaritons as described in Supplementary Section S5 (1S data is shown in Supplementary Section S3). Subtracting the lower polariton energy from the upper polariton energy yields the density-dependent Rabi splitting (normalized with zero density Rabi splitting) for both the excited 2S-exciton (Fig. 4b) and the ground state 1S-exciton (Fig. 4c). The observed Rabi splitting indeed decreases linearly in the small intensity regime as suggested by Eq.(2). In this linear regime, we can use the known linear relation between the pump intensity and the polariton density to fit the relation Eq.(2) to our measurements. This in turn allows to determine the polariton blockade radius, and yields $R_{bl}^{(2S)} = 25.1 \pm 4.0$ nm. In contrast, our analogous measurement for 1S ground state excitons in $WS_2$ reveal a significantly weaker nonlinearity and an associated blockade radius of $R_{bl}^{(1S)} = 5.8 \pm 1.1$ nm. In fact, the exciton radii have been measured experimentally[7,39] to be $a_{1S} = 1.7\ nm$ for the 1S state in $WS_2$ and $a_{2S} = 6.6\ nm$ for the 2S state in $WSe_2$, respectively. Therefore, the ratio $a_{2S}/a_{1S} = 3.9$ is indeed consistent with the enhancement $R_{bl}^{(2S)}/R_{bl}^{(1S)} = 4.3 \pm 1.1$ of the blockade radius deduced from our measurements.

The enhancement of the exciton blockade radius demonstrated in our experiments corresponds to a 16-fold enhancement of the polariton-polariton interaction. This observed growth of the nonlinearity agrees with the expected $\sim n^4$ scaling of the resulting effective polariton-polariton interaction, due to the increasing range of the exciton interaction with their principal quantum number $n$. Such a dramatic enhancement with the increasing excitonic excitation levels together with the demonstrated realization of excited-state exciton-polaritons opens up a promising approach to reaching the regime of strong polariton interactions in semiconductor microcavities.

**Methods**

**Sample fabrication.** The cavity consists of van der Waals stack of [hBN-WSe$_2$] x 3 sandwiched between a 12 period bottom SiNx/SiO$_2$ DBR, and top Poly (methyl methacrylate) (PMMA) layer and silver mirror as shown in Fig. 1a. The DBR was grown by plasma-enhanced chemical vapor deposition (PECVD) on a silicon substrate using a combination of nitrous oxide, silane and ammonia at a temperature of 350°C. Each SiO$_2$ and SiNx layer is 113 nm, 82.5 nm, with refractive indices 1.46 and 2.03, respectively. The WSe$_2$ monolayers were obtained via direct exfoliation from high-quality WSe$_2$ crystal grown by vapor flux method. The hBN capping layers were obtained through exfoliation from commercially purchased hBN crystal (HQ Graphene). Monolayer WSe$_2$ and different layers of hBN were identified under an optical microscope using their different contrasts on SiO$_2$/Si substrate. The hBN/WSe$_2$/hBN/WSe$_2$/hBN/WSe$_2$/hBN stack was realized by following the poly-propylene carbonate (PPC) pick up technique[40]. The thickness of each layer can be found in Supplementary Section S1. The completed stack was transferred onto the bottom DBR at 120 °C followed by soaking the entire sample in chloroform for 2 hours to remove PPC residue. 212 nm thick PMMA (495 A4 from Michrochem) was spin coated onto the hBN-WSe$_2$ stack to have the cavity mode in resonance with exciton energy. Finally, a 40nm thick silver top mirror was deposited using electron beam evaporation.

**Optical measurements.** The angle-dependent reflection measurements were taken using Fourier space imaging technique. A super continuum pulsed light source (NKT Photoincs) passing through a long pass filter at 650 nm (550 nm) and a short pass filter at 750nm (650 nm) was used to selectively excite the 2S (1S) exciton state in WSe$_2$ (WS$_2$). The pulse repetition rate is 40 MHz and pulse duration is 20 ps. The setup is coupled with Princeton Instruments monochromator with a PIXIS: 256 EMCCD camera.

Reference

1. Vladimirova, M. *et al.* Polariton-polariton interaction constants in microcavities. *Phys. Rev. B* **82**, 75301 (2010).

2. Ferrier, L. *et al.* Interactions in Confined Polariton Condensates. *Phys. Rev. Lett.* **106**, 126401 (2011).

3. Sun, Y. *et al.* Direct measurement of polariton–polariton interaction strength. *Nat. Phys.* **13**, 870–875 (2017).

4. Muñoz-Matutano, G. *et al.* Emergence of quantum correlations from interacting fibre-cavity polaritons. *Nat. Mater.* **18**, 213–218 (2019).

5. Delteil, A. *et al.* Towards polariton blockade of confined exciton–polaritons. *Nat. Mater.* **18**, 219–222 (2019).

6. Chernikov, A. *et al.* Exciton binding energy and nonhydrogenic Rydberg series in monolayer $WS_2$. *Phys. Rev. Lett.* **113**, 76802 (2014).

7. Stier, A. V *et al.* Magnetooptics of Exciton Rydberg States in a Monolayer Semiconductor. *Phys. Rev. Lett.* **120**, 57405 (2018).

8. Ye, Z. *et al.* Probing excitonic dark states in single-layer tungsten disulphide. *Nature* **513**, 214–218 (2014).

9. He, K. *et al.* Tightly bound excitons in monolayer $WSe_2$. *Phys. Rev. Lett.* **113**, 26803 (2014).

10. Manca, M. *et al.* Enabling valley selective exciton scattering in monolayer $WSe_2$ through upconversion. *Nat. Comm.* **8**, 14927 (2017).

11. Balili, R., Hartwell, V., Snoke, D., Pfeiffer, L. & West, K. Bose-Einstein condensation of microcavity polaritons in a trap. *Science* **316**, 1007–1010 (2007).

12. Kasprzak, J. *et al.* Bose-Einstein condensation of exciton polaritons. *Nature* **443**, 409–414 (2006).

13. Deng, H., Weihs, G., Santori, C., Bloch, J. & Yamamoto, Y. Condensation of Semiconductor Microcavity Exciton Polaritons. *Science* **298**, 199–202 (2002).

14. Amo, A. *et al.* Superfluidity of polaritons in semiconductor microcavities. *Nat. Phys.* **5**, 805–810 (2009).

15. Amo, A. *et al.* Polariton Superfluids Reveal Quantum Hydrodynamic Solitons. *Science* **332**, 1167–1170 (2015).

16. Lagoudakis, K. G. *et al.* Quantized vortices in an exciton-polariton condensate. *Nat. Phys.* **4**, 706–710 (2008).

17. Berloff, N. G. *et al.* Realizing the classical XY Hamiltonian in polariton simulators. *Nat. Mater.* **16**, 1120–1126 (2017).

18. Dreismann, A. *et al.* A sub-femtojoule electrical spin-switch based on optically trapped polariton condensates. *Nat. Mater.* **15**, 1074–1078 (2016).

19. Ballarini, D. *et al.* All-optical polariton transistor. *Nat. Comm.* **4**, 1778 (2013).

20. Sturm, C. *et al.* All-optical phase modulation in a cavity-polariton Mach–Zehnder interferometer. *Nat. Comm.* **5**, 3278 (2014).

21. Tassone, F. & Yamamoto, Y. Exciton-exciton scattering dynamics in a semiconductor microcavity and stimulated scattering into polaritons. *Phys. Rev. B* **59**, 10830–10842 (1999).

22. Liu, X. *et al.* Strong light–matter coupling in two-dimensional atomic crystals. *Nat. Photon.* **9**, 30–34 (2014).

23. Dufferwiel, S. *et al.* Exciton–polaritons in van der Waals heterostructures embedded in tunable microcavities. *Nat. Comm.* **6**, 8579 (2015).

24. Sun, Z. *et al.* Optical control of room-temperature valley polaritons. *Nat. Photon.* **11**, 491–496 (2017).

25. Dufferwiel, S. *et al.* Valley-addressable polaritons in atomically thin semiconductors. *Nat. Photon.* **11**, 497–501 (2017).

26. Chen, Y.-J., Cain, J. D., Stanev, T. K., Dravid, V. P. & Stern, N. P. Valley-polarized exciton–polaritons in a monolayer semiconductor. *Nat. Photon.* **11**, 431 (2017).

27. Barachati, F. *et al.* Interacting polariton fluids in a monolayer of tungsten disulfide. *Nat. Nanotech.* **13**, 906–909 (2018).

28. Wild, D. S., Shahmoon, E., Yelin, S. F. & Lukin, M. D. Quantum Nonlinear Optics in Atomically Thin Materials. *Phys. Rev. Lett.* **121**, 123606 (2018).

29. Walther, V., Johne, R. & Pohl, T. Giant optical nonlinearities from Rydberg excitons in semiconductor microcavities. *Nat. Comm.* **9**, 1309 (2018).

30. Kazimierczuk, T., Fröhlich, D., Scheel, S., Stolz, H. & Bayer, M. Giant Rydberg excitons in the copper oxide $Cu_2O$. *Nature* **514**, 343–347 (2014).

31. Miller, R. C., Kleinman, D. A., Tsang, W. T. & Gossard, A. C. Observation of the excited level of excitons in GaAs quantum wells. *Phys. Rev. B* **24**, 1134–1136 (1981).

32. Gammon, D., Snow, E. S., Shanabrook, B. V, Katzer, D. S. & Park, D. Fine Structure Splitting in the Optical Spectra of Single GaAs Quantum Dots. *Phys. Rev. Lett.* **76**, 3005–3008 (1996).

33. Miyata, A. *et al.* Direct measurement of the exciton binding energy and effective masses for charge carriers in organic-inorganic tri-halide perovskites. *Nat. Phys.* **11**, 582–587 (2015).

34. Galkowski, K. *et al.* Determination of the exciton binding energy and effective masses for methylammonium and formamidinium lead tri-halide perovskite semiconductors. *Energy Environ. Sci.* **9**, 962–970 (2016).

35. Piętka, B. *et al.* 2S Exciton-Polariton Revealed in an External Magnetic Field. *Phys. Rev. B* **96**, 81402 (2017).

36. Bao, W. *et al.* Observation of Rydberg exciton polaritons and their condensate in a perovskite cavity. *Proc. Natl. Acad. Sci.* **116**, 20274–20279 (2019).

37. Schneider, C., Glazov, M. M., Korn, T., Höfling, S. & Urbaszek, B. Two-dimensional semiconductors in the regime of strong light-matter coupling. *Nat. Comm.* **9**, 2695 (2018).

38. Ciuti, C., Schwendimann, P. & Quattropani, A. Theory of polariton parametric interactions in semiconductor microcavities. *Semicond. Sci. Technol.* **18**, S279–S293 (2003).

39. Shahnazaryan, V., Iorsh, I., Shelykh, I. A. & Kyriienko, O. Exciton-exciton interaction in transition-metal dichalcogenide monolayers. *Phys. Rev. B* **96**, 115409 (2017).

40. Wang, L. *et al.* One-Dimensional Electrical Contact to a Two-Dimensional Material. *Science* **342**, 614–617 (2013).

**Acknowledgements:** We acknowledge support from the National Science Foundation through the EFRI-2DARE program (EFMA-1542863), MRSEC program DMR - 420634 and the ARO MURI program (W911NF-17-1-0312). LW acknowledges support by the Alexander von Humboldt foundation. AR gratefully acknowledges support through the Laboratory Directed Research and Development Program of Lawrence Berkeley National Laboratory under U.S. Department of Energy Contract No. DE-AC02-05CH11231. The authors also acknowledge the use of the Nanofabrication Facility at the CUNY Advanced Science Research Center for the fabrication of the devices.

**Author Contributions:**
V.M.M., T.F.H., J.G conceived the experiments. J.G. fabricated the devices. D.R., J.C.H. grew the bulk $WSe_2$ crystal. J.G., L.W., performed the measurements. J.G. performed data analysis. T.P., S.K.C., V.W. did the theoretical modeling. All authors contributed to write the manuscript and discuss the results.

**Competing Interests:** The authors declare that they have no competing financial interests.

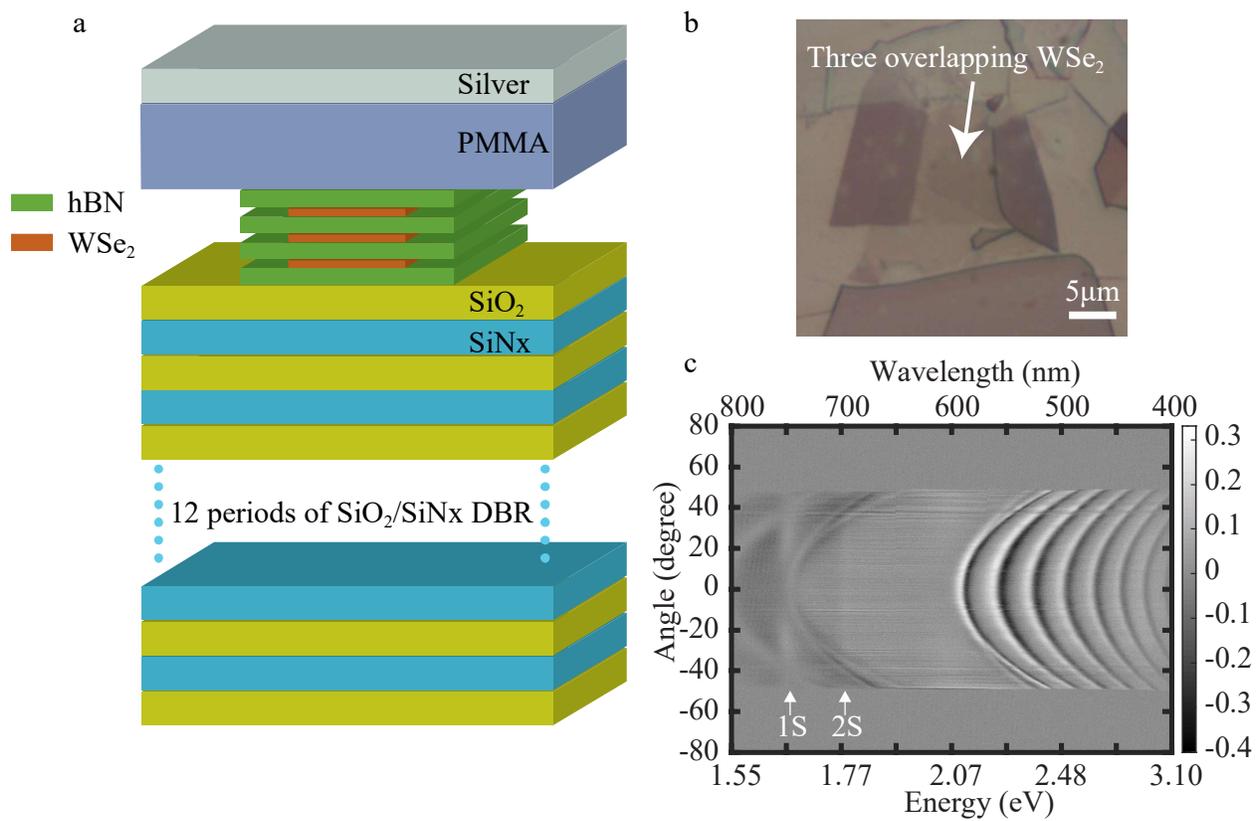

**Figure 1 | Device structure. a**, Schematic illustration of the sample structure. **b**, Optical image after the multilayer hBN-WSe$_2$ stacking was transferred on top of the DBR. The area where three WSe$_2$ layers overlap is marked with the white arrow. **c**, Angle resolved differential reflection of multilayer structure on DBR at room temperature. Absorption features can be observed at the energies of the 1S and 2S exciton and are marked by the white arrows.

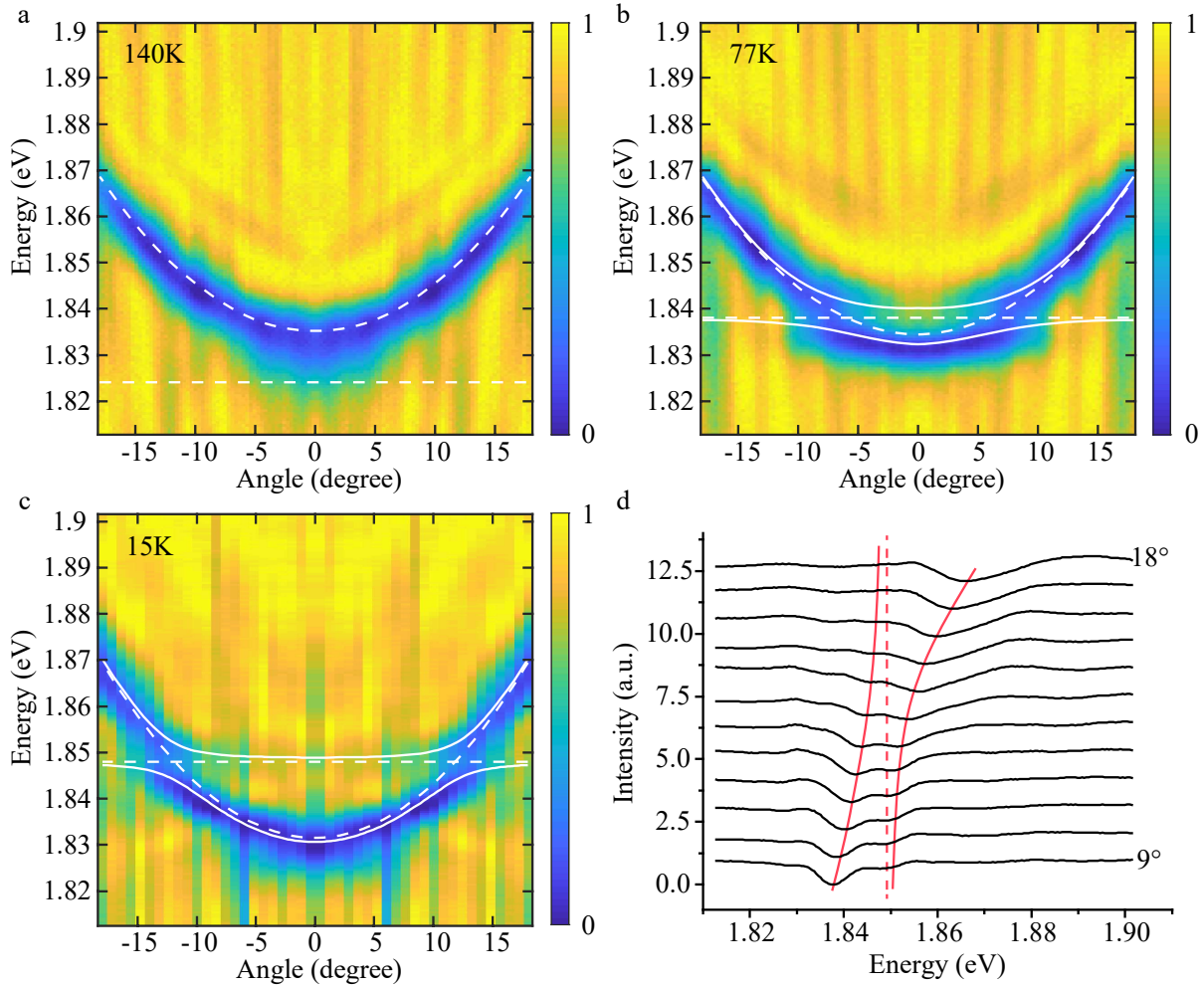

**Figure 2 | Temperature dependent cavity coupling to excited-state excitons. a-c**, Angle resolved reflection at different temperatures of 140K, 77K, 15K. The dashed lines indicate the bare cavity and exciton dispersions, while the solid lines correspond to the resulting dispersion relations of the two polaritons induced by the strong coupling between the excitons and the cavity photons. **d**, Line cut from the 15K data between 9° to 18°. Those data have y axis offset for clarity. A clear anti-crossing is observed (red trend lines). The extracted Rabi splitting is 7.7 meV. The dashed line shows the 2S exciton energy at 1.848 eV.

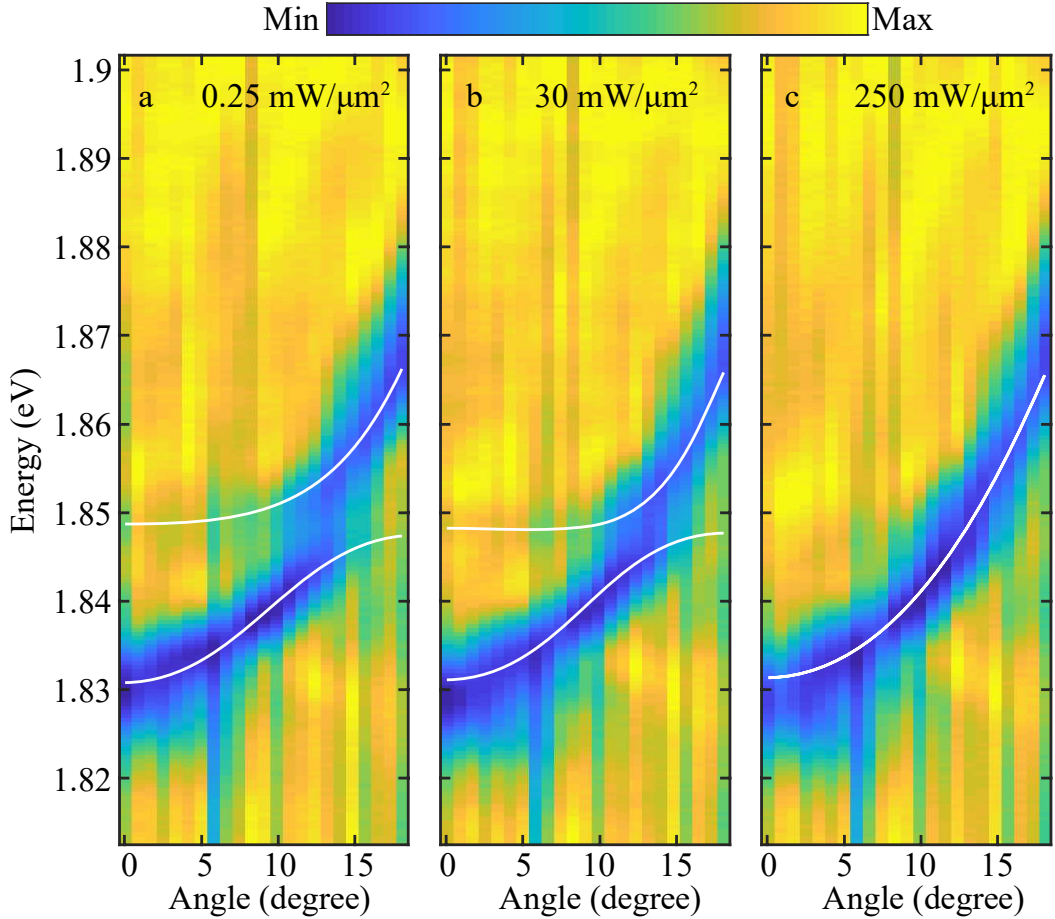

**Figure 3 | Intensity dependent cavity reflection.** a-c, Angle resolved reflection for different incident intensities of 0.25 mW/μm² (**a**), 30 mW/μm² (**b**) and 250 mW/μm² (**c**). Higher intensities imply a higher density of polaritons, which eventually leads to an excitation blockade induced by the strong interaction between excitons. This polariton blockade reduces the effective coupling to the cavity, observed as a decreasing splitting of the two polariton branches in our experiments.

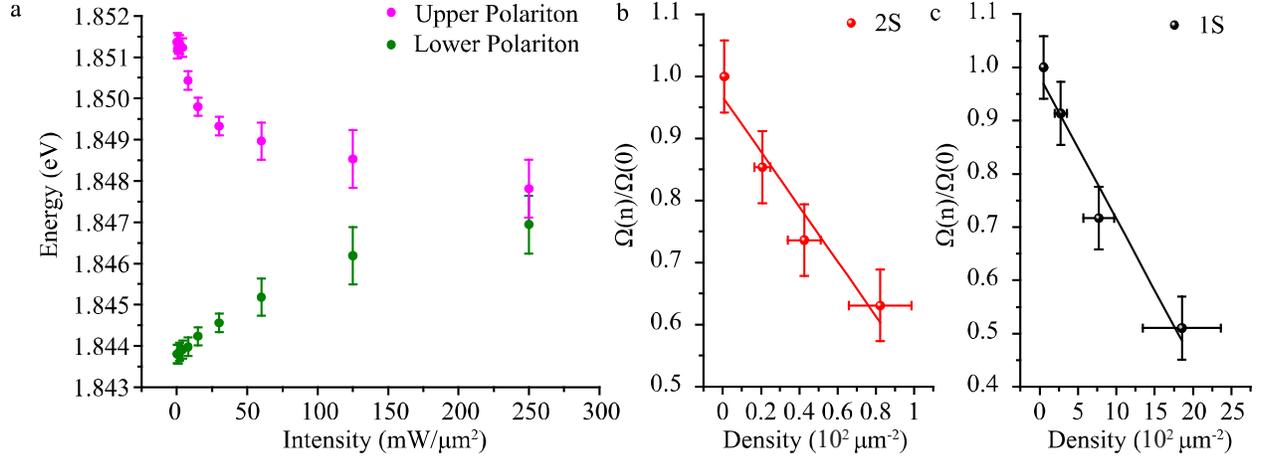

**Figure 4 | Interaction induced nonlinear cavity response. a**, Energies of the upper polariton and lower polariton under different injected intensity at the Rabi splitting angle where exciton and photon component have the same fraction. **b**,**c** Normalized Rabi splitting as a function of the polariton density for 2S (**b**) and 1S (**c**). The solid lines show fits of Eq.(2) to our low-density data, yielding blockade radii of 25.1 ± 4.0 nm and 5.8 ± 1.1 nm for the 2S- and 1S-exciton, respectively.